\documentclass[runningheads]{svmult}

\usepackage{makeidx}   
\usepackage{graphicx}  
\usepackage{subeqnar}  
\usepackage{multicol}  
\usepackage{physprbb}  
\makeindex             

\begin{document}
\title*{Galaxy formation and evolution since z= 1}
%
%
%
%
\titlerunning{galaxy formation at z$\le$ 1}
%
\author{Fran\c{c}ois Hammer\inst{1}, Hector Flores\inst{1}, YanChun Liang\inst{1}, XianZhong Zheng\inst{1}
\and David Elbaz\inst{2}
\and Catherine Cesarsky\inst{3}}

\authorrunning{Fran\c{c}ois Hammer et al.}
%
%
\institute{GEPI, Observatoire de Paris, 92195 Meudon Principal, France
\and CEA, Saclay-SAp, Ormes des Merisiers, 91191 Gif sur
Yvette, France
\and ESO, Karl-Schwarzschild Strase 2, D85748 Garching bei Munchen, Germany}

\maketitle              

\begin{abstract}
Determination of the star formation rate can be done using mid-IR photometry or
Balmer line luminosity after a proper correction for extinction effects. Both methods show
 convergent results while those based on UV or on [OII]$\lambda$3727
  luminosities underestimate the SFR by factors ranging from 5 to 40 for starbursts
and for luminous IR galaxies, respectively. Most of the evolution of the cosmic star
formation density is related to the evolution of luminous compact galaxies and to
luminous IR galaxies. Because they were metal deficient and were forming stars at very high 
rates (40 to 100 $M_{\odot}yr^{-1}$), it is probable that these (massive) galaxies were actively
 forming the bulk of their stellar/metal content at z $\le$ 1.      
\end{abstract}

\section{Introduction}
Determining how galaxies formed and grew is one of
the outstanding problems of modern astrophysics.
 The epoch at which half the stellar mass
was formed has been estimated to be z=1-1.5 from simple integrations of the global star
formation history of the Universe ([1]; [2]). Similar results have been found 
by estimating the growth of galaxies by calculating the stellar mass from SEDs
with broad wavelength coverage ([3]; [4]; [5]).  However, this method depends strongly
on the estimated M/L of individual galaxies which is sensitive to the
assumptions about their star formation history, dust distribution,
metallicity, and IMF -- all of which are usually poorly constrained.
Many efforts have been made to study galaxies at redshifts higher than 1, in
attempts to track the formation of half the present day stellar mass. However the situation at
z$>$ 1 is difficult, because of their faintness and because most of the important lines for diagnostics are redshifted to
the near IR. Here we choose to investigate the properties of z $\le$ 1 galaxies, for
which the current generation of telescopes is able to provide accurate measurements of
their properties (star formation rates, masses, metal abundances). In a
cosmological model with $H_{0}$ =70km $s^{-1}$ $Mpc^{-3}$, $\Omega_{\Lambda}$=0.7 and
$\Omega_{M}$=0.3, z=1 corresponds to 58\% of the Universe age. At z$>$ 0.4,
the emergence of two galaxy populations have been reported, namely the luminous infrared
galaxies (LIRGs, [1]; [6]) and the luminous compact
galaxies (LCGs, [7]; [8]). These two populations are responsible
of most of the evolution of the IR and UV luminosity density evolutions, and then of a
significant fraction of the cosmic star formation density at z$\sim$ 1.   

In the following, we present a summary of results based on follow-up studies of the Canada France
Redshift Survey (CFRS) using the Very Large Telescope (VLT), the Infrared Space Observatory (ISO)
and the Hubble Space Telescope (HST).

\section{Estimating extinction and SFR at z$\sim$ 1} 

A prerequisite for estimating star formation from emission is to properly estimate the
extinction. $H\alpha$ luminosity is one of the best indicator of the instantaneous SFR.  
However low resolution spectroscopy (R $<$ 500) often produces misleading results [9]. Only good S/N 
spectroscopy with moderate spectral resolution (R$>$600) 
allows a proper estimate of the extinction from the $H\alpha/H\beta$ ratio after
accounting for underlying stellar absorption. SFRs can be otherwise underestimated
or overestimated by factors reaching 10, even if one accounts for an {\it ad hoc} extinction 
correction. These effects are prominent for a large fraction of evolved massive galaxies especially those experiencing
successive bursts (A and F stars dominating their absorption spectra). Further estimates of the cosmic star formation
density at all redshifts mandatorily requires moderate resolution spectroscopy to avoid severe and uncontrolled biases.  

Liang et al ([9]) have given an obvious warning for the studies based
on low resolution spectroscopy aimed at measuring individual galaxy properties 
(gas chemical abundances, interstellar extinction, 
stellar population, ages as well as star formation rates and history),
particularly for dusty spiral galaxies. For example, it has been
shown ([10]) that 1/4 of the Lilly et al sample ([11]) of z $\sim$ 0.7
galaxies was made of LIRGs which are dust enshrouded systems (average $A_{V}$= 2.4).
Assuming a constant extinction of $A_V$=1 for these LIRGs, leads to underestimate
their [OII]$\lambda$3727/$H\beta$ ratio, providing an overestimate of their 
 metallicity by 0.3 to 0.5 dex.
 
\begin{figure}[]
\begin{center}
\includegraphics[width=1.0\textwidth]{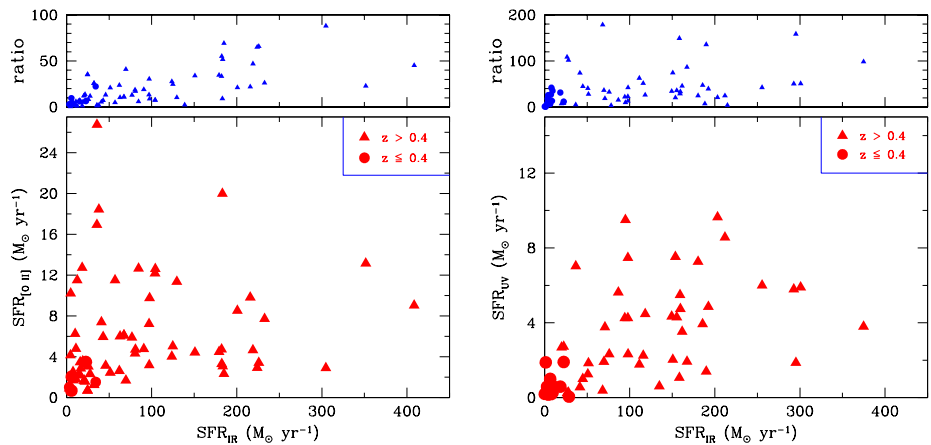}
\end{center}
\caption[]{Comparison of SFRs calculated using the formalism of [13].
({\it Left panel}): [OII]$\lambda$3727 estimates are compared to IR estimates for a sample of 70
galaxies from the CFRS and Marano fields (see [10]). [OII]$\lambda$3727 
estimates have been done using the formulae from [7] which have been
adequately multiplied by 2.46 to account for the fact that they have used a different IMF
than [13].  Upper left panel provides
the $SFR_{IR}/SFR_{[OII]}$ ratio. Median values of the ratio are 5 and 22 for starbursts (SFR $<$
40 $M_{\odot} yr^{-1}$ and LIRGs respectively. ({\it Right panel}): same comparison for
 UV (2800\AA~) luminosity estimated for 61 CFRS galaxies. Median values of the 
$SFR_{IR}/SFR_{UV}$ ratio are 13 and 36 for starbursts (SFR $<$
40 $M_{\odot} yr^{-1}$ and LIRGs, respectively.}
\label{eps1}
\end{figure}

ISOCAM observations give us a unique opportunity to test the validity of our SFR
estimates. It has been shown ([6], see also David Elbaz's contribution) that the
mid-IR and radio luminosities correlate well up to z=1. Bolometric IR measurements derived
from mid-IR are validated, unless if both the radio-FIR and the MIR-FIR correlations
are unvalid at high redshifts. Mid-IR photometry provides an unique tool to estimate
properly the SFR of LIRGs (defined as $L_{IR} >$ 2 $10^{11}$ $L_{\odot}$) up to z=1.2, 
and of starbursts ($L_{IR} <$ 2 $10^{11}$ $L_{\odot}$) up to z=0.4. For a sample of 16
ISO galaxies at 0 $<$ z $<$ 1, 
 SFRs have been derived from the extinction corrected $H\alpha$ luminosities ([12]). These
values agree within a factor 2 with mid-IR estimates (median value of
$SFR_{IR}/SFR_{H\alpha}$=1.05). Using moderate spectral
resolution (R = 1200) with good S/N of 90 ISO galaxies (0.2 $<$ z $<$ 1), Liang et al [10] have shown
that the extinction can be properly derived from the $H\beta/H\gamma$ ratio. 
Such measurements can be performed for all galaxies up to z=1 within an accuracy of 0.6 mag for
$A_{V}$ values, and then for SFR values. 

It has been argued ([13]) that the [OII]$\lambda$3727 luminosity can provide an
efficient way to derive SFRs of high redshift galaxies using spectrographs in the visible
range. However preliminary results from [14] and [15] have
shown that the correlation between $H\alpha$ and [OII]$\lambda$3727 is rather poor, because of
extinction effects and also because it considerably depends on the galaxy metallicity, luminosity and
spectrophotometric type. How [OII]$\lambda$3727 estimates compare to IR (or
$H\alpha$) estimates of the SFR ? Figure 1 shows that the former strongly underestimates
the SFR by factor 5 for starbursts and factor 22 for LIRGs. A similar result is found
by comparing SFR estimates from UV to those from IR, and has been already discussed by
[8] for a sample of distant LCGs (average SFR= 40 $M_{\odot} yr^{-1}$). The fact
that UV fluxes provide even lower SFR values than [OII]$\lambda$3727 fluxes might be related to the expected increase of
[OII]$\lambda$3727/$H\beta$ ratio with decreasing metal abundance (high z systems being expected to be less metal abundant than local
ones).

\section{Were all massive galaxies already formed at z=1 ?}

It has been claimed ([16]) that the stellar mass built up of massive
spirals was mostly done at z$\sim$ 1. This claim was based on the very low values
of the $SFR/M_{stellar}$ for CFRS galaxies, yielding very low stellar
mass increases (ranging from 1 to 10\%) for massive galaxies ($M_{stellar}$ $>$
$10^{11}M_{\odot}$) since z=1. If true, this result would imply that most of the star
formation at z$<$ 1 occurred in dwarf galaxies. However, present day dwarves contain
only a marginal fraction of the stars or metals, and the Brinchman and Ellis 's result ([16]) is
 somewhat at odd with integrations of the star formation history and with stellar mass
 density at different lookback times. We notice that SFRs calculated by [16] were
  based on the formulae from [7], and provide values 54 and 12 times
 lower than our estimates for LIRGs and starbursts, respectively. This potentially leads to a severe
 revision of the conclusions from [16].
 
\begin{figure}[]
\begin{center}
\includegraphics[width=1.0\textwidth]{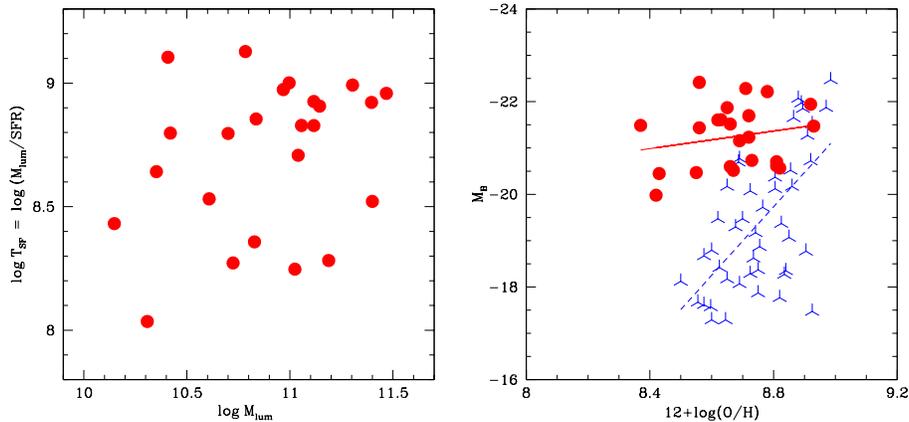}
\end{center}
\caption[]{({\it Left panel}): $M_{stellar}/SFR$ against $M_{stellar}$ for distant LIRGs (from [17]).
 Because of their very high SFRs, LIRGs can build up of significant part of 
 their masses in relatively
 short times.({\it Right panel}): $M_{B}$ versus O/H abundance ratio of 
 distant LIRGs (from [10], full dots) compared to
 those of local spirals ( from [18] and [15], skeletal symbols). Full line and dotted lines report
 the regression law for LIRGs and local spirals, respectively. At z$\sim$ 0.75, LIRGs have log(O/H) values
  0.3 dex lower than those of local spirals.}
\label{eps1}
\end{figure}

 A pioneering study of 14 distant LCGs ($M_{B}<$ -20; $r_{half}$ $\le$ 3.5 $h_{70}^{-1}$ kpc) have revealed that they are
 dust enshrouded starbursts superimposed to an old and evolved stellar population. It has lead us ([8]) 
 to suggest that LCGs are the progenitors of present-day spiral bulges, prior to the star
 formation in the disk component. Because LCGs correspond to 23\% of the luminous galaxy population at z
 $\sim$ 0.75, they could be the progenitors of a similar fraction of present-day luminous spirals.  
 
 Recently, important progress has been made in our understanding of distant LIRGS 
 ([10]; [17]). They are forming rapidly 
 new generations of stars and possess stellar masses comparable or slightly smaller than that of
 present-day massive spirals
  (Figure 2; [17]). Galaxies experiencing such strong star
  forming events could form
  a significant fraction of their masses during the last 8 Gyr. 
  Their metal content have been found to be on average half that
   of today massive spirals (Figure 2; [10]): at such rates of star/metal 
   production, they can  be the progenitors of  massive spirals within timescales 
   much smaller than a Hubble time.
   Indeed LIRG morphologies are intimately related to giant
    disks (Figure 3; [17]), and, incidentally include a significant fraction 
    of compact galaxies. Because LIRGs and LCGs are much more frequent at z$>$
    0.4 than today, a substantial fraction of massive galaxies could be actively forming
    their stellar content at z $<$ 1. It does not necessarily contradict previous
    studies (e.g. [19]) which have shown that the number density of
    large spirals ($r_{disk}$ $>$ 2.8 $h_{70}^{-1}$ kpc) was similar at z=0.75 than
    today. In fact we realize that one third of the large disks in the sample of [19] were
    indeed LIRGs, strengthening our prediction that many large
    spirals were actively forming the bulk of their stellar content at z $<$ 1.

\begin{figure}[]
\begin{center}
\includegraphics[width=1.0\textwidth]{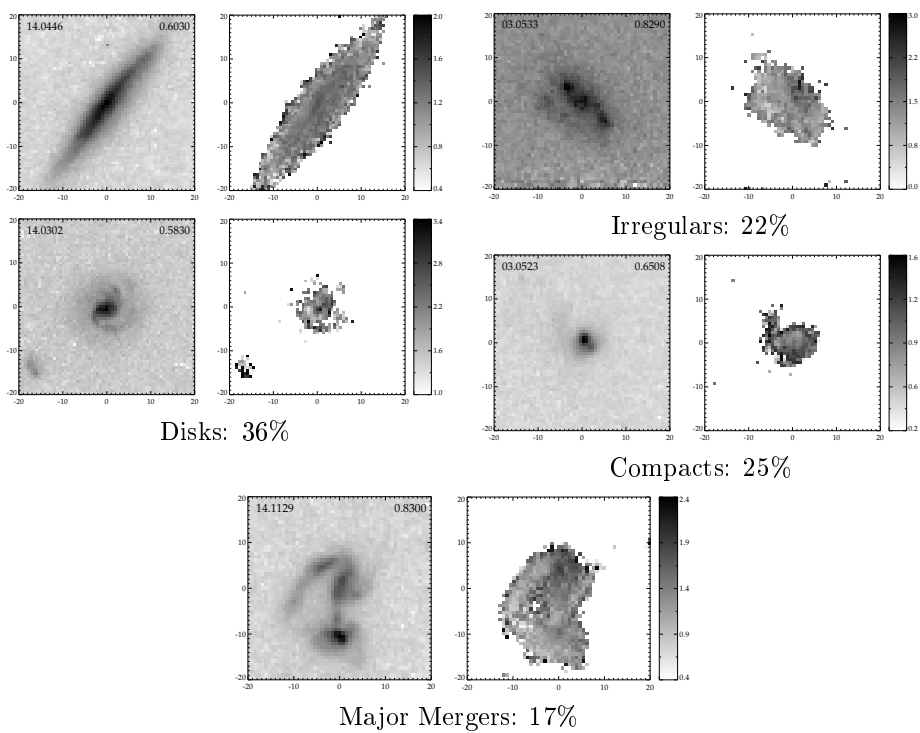}
\end{center}
\caption[]{Morphological classifications using color maps from a study of 36 LIRGs by
[17]. Using color map and deconvolution software (GIM2D), it has succeeded
to classify all the galaxies (fractions of each type are given). For each galaxy, the left stamp (40x40 $kpc^{2}$ area) 
shows the F814W WFPC2 image, and the right stamp shows the color map (F606W-F814W) following
the prescription of Zheng et al ([17]). Color map have been derived after a very accurate
alignment of V and I images down to 0.015 arcsec, which is a prerequisite for galaxies
with some irregularities in their morphologies. Notice the fact that most distant 
galaxies show small blue and very red regions, interpreted as HII regions and dusty
enshrouded regions, respectively. Notice also the very blue color in the center of
compact galaxies.}
\label{eps1}
\end{figure}

    Extragalactic studies might appear unaccurate and affected by numerous biases and
    uncertainties when compared to WMAP results (see Licia Verde's contribution): they 
     have been nicknamed "dirty physics" during the first talk of the Conference. Indeed,
     the extensive use of photometric redshifts or of low resolution spectroscopy of z$>$ 1 galaxies
     can cast some doubts about our knowledge of their properties.   
    On the other hand, physical parameters of z$\le$ 1 galaxies can be determined rather
    accurately, because these sources are enough bright to be at the reach of current
    telescopes. This requires to apply methodologies (for example high S/N and medium resolution
    spectroscopy) which have been successful for
    analysing local galaxy properties. In this paper, we also systematically compare 
     two independent estimates of each individual parameter: SFR calculations have been done by
    comparing IR and Balmer line luminosities and stellar production in LIRGs and in LCGs 
    have been compared to their metal content relative to that of local spirals. 
    
    We intend
    to pursue our analyses towards the general population of field galaxies up to z=1,
    including by studying their morphologies, metal content and SFRs (Hammer et al, 2004, in preparation). We
    believe that establishing a new classification sequence for all galaxies at z $<$ 1
    is at the reach of the present generation of telescopes. An important goal will be to study their
    dynamics using the multi-integral field unit mode of FLAMES/GIRAFFE at VLT. It would
    provide an important test for the merging scenario, as well as a solid estimate of the
    evolution of the Tully-Fisher relation up to z=1. Dynamical masses of galaxies could be then compared to
    their stellar masses.

%

\end{document}